\documentclass[superscriptaddress,amsmath,amssymb,aps,epsf,prb,twocolumn]{revtex4-2}
\usepackage[T1]{fontenc}
\usepackage{bm}
\usepackage{amsmath}
\usepackage{amsfonts}
\usepackage{amssymb}
\usepackage[dvipsnames]{xcolor}
\usepackage{hyperref}
\usepackage{graphicx}

\usepackage[normalem]{ulem}

\usepackage{braket}

\begin{document}
	\title{Inelastic scattering and transient localization from coupling to two-level systems}
	\author{Hadi Rammal}
	\affiliation{Universit\'e Grenoble Alpes, CNRS, Grenoble INP, Institut N\'eel, 38000 Grenoble, France}
	\author{Sergio Ciuchi}
	\affiliation{Dipartimento di Scienze Fisiche e Chimiche, Universit\'a dell’Aquila, Coppito-L’Aquila, Italy}
	\author{Simone Fratini}
	\affiliation{Universit\'e Grenoble Alpes, CNRS, Grenoble INP, Institut N\'eel, 38000 Grenoble, France}

\begin{abstract}
We consider an electron interacting locally with two-level systems (TLSs) as an archetypal model for charge transport in the presence of inelastic scatterers. To assess the importance of quantum effects in the optical and d.c. conductivity  we solve  the model numerically without approximations using the finite temperature Lanczos method (FTLM), and 
compare the results with Dynamical Mean Field Theory (DMFT).
In the slow fluctuation limit, the coupling to the TLSs causes transient localization of the carriers analogous to the one recently found in the electron-boson scattering problem,  featuring enhanced resistivities and displaced Drude peaks. Fast inelastic scatterers  suppress localization, restoring a more conventional regime where  transport and optical properties are governed by independent scattering events.
\end{abstract}

\maketitle

\noindent

\section{Introduction} 
A general feature of strongly interacting electron systems is that their charge transport properties  can contradict, to varying degrees, several of the accepted assumptions that instead broadly apply in weak scattering conditions. Common experimental anomalies include resistivities that far exceed the Mott-Ioffe-Regel limit, typically found in bad semiconductors \cite{AdvMat16} and bad metals \cite{Hussey}, and resistivities that vary linearly with temperature over extended temperature windows, in strange metals \cite{PhillipsScience22}. 

Among the many available theoretical scenarios leading to anomalous transport properties,
it has been recently recognized that the scattering of electronic carriers by sufficiently slow  bosons can induce localization 
even in the absence of extrinsic disorder  \cite{Troisi06,PRBR11,AdvMat16,rammalPRL2024}, challenging  the most fundamental assumptions of Boltzmann transport theory.   This is an inherently quantum phenomenon, originating from  the same wavefunction interferences that are known to cause Anderson localization, and occurring at times shorter than the typical timescale of boson dynamics.
Such \textit{transient  localization}  generally leads to an increase in resistivity and can therefore naturally yield bad conduction in both semiconductors \cite{PRBR11,AdvMat16,rammalPRL2024} and metals \cite{SciPost,HellerPRL}.
Transient localization also has notable consequences on the frequency response of the current carriers: the same non-local quantum interferences causing the observed increase in resistivity (in diagrammatic terms, vertex corrections in the current-current correlation function) are responsible for the appearance of  finite-frequency absorption peaks. Such displaced Drude peaks are direct signatures of quantum localization, with their characteristic frequency $\omega_L$ being a direct measure of the  localization radius of the carriers' wavefunctions \cite{SciPost,rammalPRL2024}.


In this work we propose to study the fundamental problem of inelastic scattering of electronic carriers by quantum degrees of freedom
by introducing a theoretical framework that is alternative to the existing electron-boson interaction models. Specifically, we consider the problem of an electron interacting with quantum two-level systems (TLSs), and study the resulting transport properties numerically without approximations via the finite temperature Lanczos method (FTLM) \cite{FTLM_1,prelovsek,rammalPRL2024}. This method fully includes quantum corrections to transport, 
hence giving access to localization physics.

TLSs are the simplest inelastic scatterers one can think of, yet their effect on charge transport hasn't been  explored in full. We anticipate here the main result of our study: when the Rabi scale $\omega_0$ of the TLSs is slow compared to that of carrier hopping on the lattice, their inelastic scattering induces transient localization of the carriers. 
This extends the realm of this phenomenon beyond electron-boson interactions \cite{rammalPRL2024}, and questions the broad applicability of large-$N$ approaches \cite{TLSJEORGE1,TLSJEORGE2} and dynamical mean field theory (DMFT), that neglect quantum interference effects. The weak scattering semiclassical picture is safely recovered  when the TLS fluctuations are fast. i.e. at large $\omega_0$, in which case quantum corrections to transport become irrelevant.

\section{Model and methods}

We consider the following model Hamiltonian:
\begin{equation}
    H = -t \sum_{\langle i j \rangle} c_{i}^{\dagger}c_{j}  + \frac{\omega_0}{2}\sum _i \sigma^z_i 
     - g \sum _i c_{i}^{\dagger}c_{i}\sigma^x_i, 
\label{eq:TLSNodisorder}
\end{equation}
representing electrons moving on a lattice (creation and annihilation operators $c_{i}^{\dagger}$ and $c_{j}$, hopping integral $t$) and coupled with two-level systems (TLS) located at each lattice site (interaction strength $g$). These are described by  spin operators $\mathbf \sigma_i$, with an energy splitting  between the two levels set by the $\sigma^z_i$ component, determining the Rabi frequency $\omega_0$. The local interaction with the electron density causes spin flip transitions, driven here by the $\sigma^x_i$ component.   

The model, which  describes the local coupling of electronic carriers with  dynamical  spin degrees of freedom, 
can also serve as an effective model in a variety of physical situations. A typical example is the interaction with local electronic or structural degrees of freedom in glasses \cite{TLSJEORGE1,PRLIntrinsicQuantumGlasses,AndersonPhilMag72}, in which case the energies $\omega_0$ and tunneling matrix elements $g$ are considered to be randomly distributed. 
TLSs 
have been recently  considered as effective low-energy degrees of freedom causing  strange metal behavior in models with random interactions \cite{TLSJEORGE1,TLSJEORGE2}.
Moving beyond this phenomenological approach, it has been suggested that superconducting puddles effectively behave as TLSs, scattering the electronic carriers and causing strange metal transport \cite{Bashan25}. 
Finally, the model Eq. (\ref{eq:TLSNodisorder}) can  be viewed as a reduction of the Holstein model describing local electron-boson interactions, with the harmonic oscillator basis being truncated to include the ground state and the first excited state only \cite{CaponeCartaPRB06}. 

In the absence of energy splitting, $\omega_0=0$, the spin operators are  diagonal along the $x$ axis and do not fluctuate.  At any nonzero temperature,
both spin states 
are  statistically populated, and the model reduces to that of electrons moving on a lattice in the presence of static disorder, $H = -t \sum_{\langle i j \rangle} c_{i}^{\dagger}c_{j}  + \sum _i \epsilon_i c_{i}^{\dagger}c_{i}$ with the local site energies randomly distributed according to the binary distribution $P(\epsilon_i)=w\delta(\epsilon_i- g)+(1-w)\delta(\epsilon_i+ g)$. The relative probabilities become equal in the limit $T\gg \omega_0$,  ($w\to 1/2$), while at  temperatures $T\ll \omega_0$, the fraction $w$ of defects is exponentially suppressed (see Appendix \ref{app:DMFT}). 
This mapping to a disordered model
implies that Anderson localization can be realized in the TLS interaction model when $\omega_0\to 0$, and that footprints of localization  should persist for small nonzero $\omega_0$:  transient localization should arise in the  TLS interaction model at low Rabi frequencies $\omega_0$, analogous to the one observed in the Holstein model for electron-boson interactions at low boson frequencies \cite{rammalPRL2024}.

In practice, we solve Eq. (\ref{eq:TLSNodisorder})
exactly using the finite temperature Lanczos method  \cite{FTLM_1,prelovsek,rammalPRL2024} for a single electron on one-dimensional chains of up to $N_s=18$ sites. These sizes are much larger than those accessible for the Holstein model ($N_s\le 6$ \cite{rammalPRL2024}), which is a consequence of the drastic reduction of Hilbert space size implied by having only two states per site. As a result, in the regime of low Rabi frequencies $\omega_0/t\lesssim 0.5$ where the TLS fluctuations are slow and  localization effects are strongest, finite-size effects are well controlled    already by using periodic boundary conditions. For larger values of $\omega_0$ we  resort instead to twisted boundary condition averaging \cite{rammalPRL2024} in order to further reduce finite-size effects when there is no wavefunction localization (see below).

We additionally solve the problem  using dynamical mean-field theory (DMFT, details in Appendix \ref{app:DMFT}). As in the electron-boson problem \cite{depolarone}, DMFT yields an analytical solution 
for the self-energy of the TLS-interaction problem on the real frequency axis, which holds directly in the thermodynamic limit and is free from the common uncertainties related to numerical or approximate impurity solvers. DMFT provides a very accurate description of charge transport at the semiclassical level, but by construction it does not include vertex corrections in the evaluation of the current-current correlation function. This means that localization processes are totally left out: the comparison between the FTLM and DMFT results  therefore allows us to quantitatively assess the importance of current vertex corrections responsible for localization.

\section{Transient localization in the regime of slow TLS fluctuations}
\subsection{Optical conductivity and transport}

\begin{figure}[h]
        \includegraphics[width=9cm]{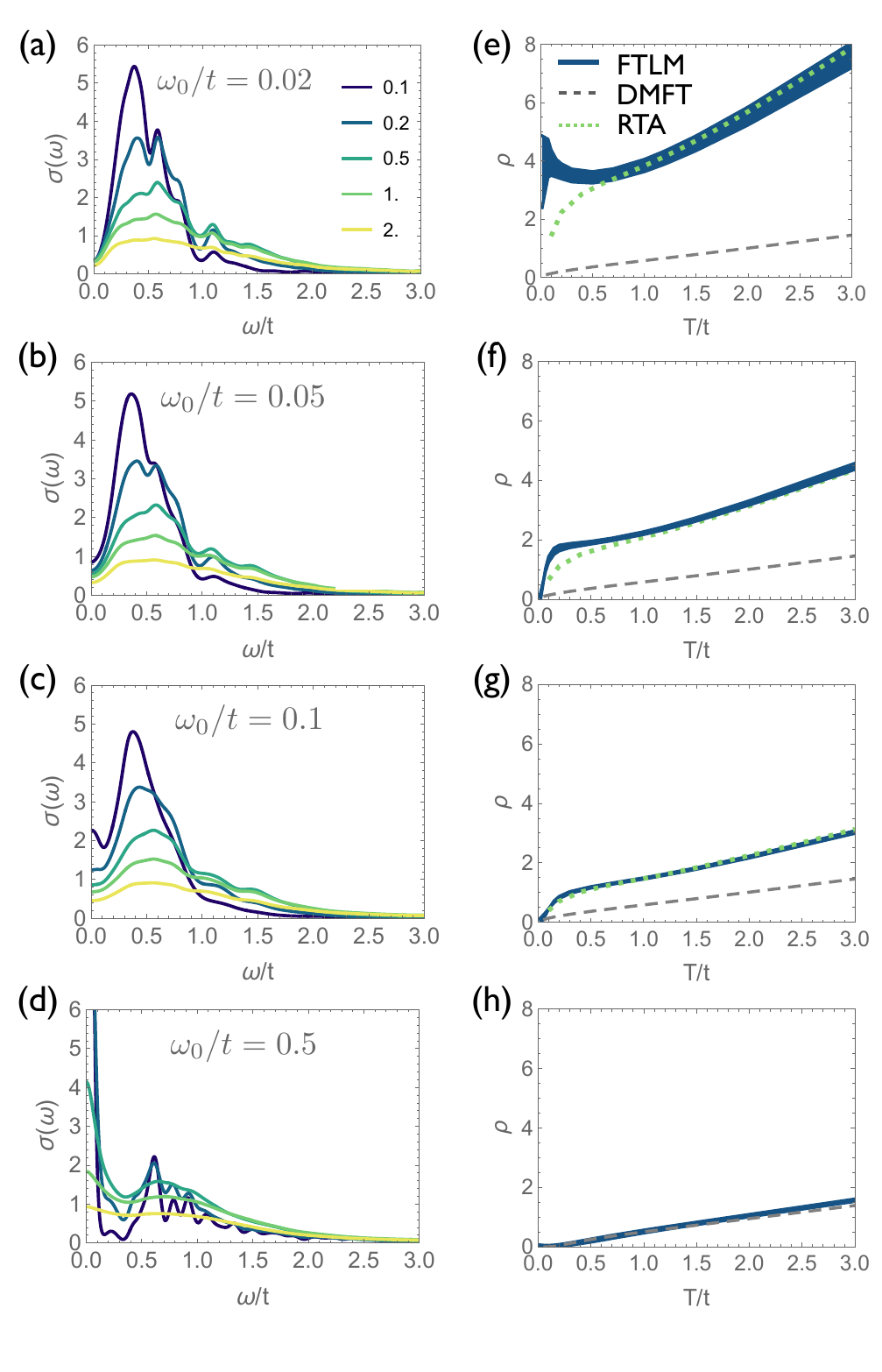}
\caption{(a-d) Optical conductivity per particle calculated from FTLM for an electron-TLS coupling $g/t=1$ on a chain of length $N_s=16$ with periodic boundary conditions, with $M=1200$ Lanczos iterations and $N_r=100$  realizations. Different panels correspond to different values of $\omega_0$ as indicated. Temperature legends are indicated in panel (a).  A Gaussian filter of width $\delta=0.04t$ has been applied. (e-h) Corresponding d.c. resistivities as a function of temperature $T$. The error bands correspond to a 5-fold variation of the filter width in the interval $0.003<\delta<0.015$ (panels e,f) and $0.008<\delta<0.04$ (panels g,h).  The dashed and dotted lines are respectively the DMFT and RTA results (see text). } 
    \label{fig:spectra-vsT}
\end{figure}

Fig. \ref{fig:spectra-vsT} shows the frequency-dependent optical conductivity per carrier, $\sigma(\omega)$ at different temperatures (labels, left panels) and resulting T-dependent resistivity, obtained as $\rho=1/\sigma(\omega\to 0)$, for $g=t=1$ and different values of the TLS dynamical scale $\omega_0$ (labels, right panels). All the data are converged with respect to the system size (here we use $N_s=16$) as well as to the number of realizations ($N_r=100$) and iterations ($M=1200$) of the Lanczos algorithm.

For sufficiently low $\omega_0$, Fig. \ref{fig:spectra-vsT}(a-c), the optical absorption at all temperatures shown exhibits a marked finite-frequency peak at a frequency $\omega_L\simeq 0.5t$. 
As anticipated previously,  this peak is indicative of quantum localization of the electronic wavefunction, with a  localization length $L$ that can be estimated from the relation $L/a=\sqrt{2t/\omega_L}$ \cite{SciPost,rammalPRL2024} and corresponds here to 2-3 lattice spacings. 
This short localization length  is responsible for the good convergence of the numerical results at the cluster sizes used in this study. The intensity of the "displaced Drude peak" (DDP) in the optical conductivity is progressively reduced with increasing temperature, while its shape broadens.

As shown  in panels (e,f,g), when a DDP is present the  resistivity is always larger than the semiclassical value calculated from DMFT (dashed lines), also indicative of sizable localization corrections. Non-local localization effects are most striking at very low Rabi frequencies, $\omega_0/t=0.02$, where the resistivity is not only much larger than the semiclassical estimate, but it additionally shows an upturn upon lowering the temperature below $T\sim t$. 
This weakly activated behavior is rapidly lost upon increasing $\omega_0$ (panels (f,g)), being replaced instead by a weakly increasing resistivity with a finite $T=0$ intercept.

Regardless of the value of $\omega_0$, the resistivity evaluated from both FTLM and DMFT eventually drops exponentially when $T\lesssim \omega_0$. It must vanish at $T=0$ all the TLSs become aligned on the lowest-energy state and become therefore unable to scatter the carriers (cf. Appendix \ref{app:DMFT}). In the opposite high-temperature limit, $T\gtrsim t$, the resistivity increases linearly with temperature. This is the normal behavior expected in  interacting systems when the temperature exceeds the total width of the excitation spectrum, which can be understood directly from the Kubo formula \cite{PRBStatisticaltheory,auerbach,spin-Holstein}.

\subsection{Crossovers and two-peak coexistence}

At the largest Rabi frequency explored ($\omega_0/t=0.5$, panel (h)) the resistivity becomes almost indistinguishable from the DMFT result, indicating that quantum corrections have been essentially washed out.

To substantiate this progressive disappearance of localization effects, let us take a closer look at the optical conductivity spectra shown in panels (a-d). Increasing the Rabi frequency $\omega_0$ reveals two systematic trends. First, the exponential suppression of the resistivity in the low-temperature regime $T\lesssim \omega_0$ (right panels) is  associated with the emergence of a narrow $\omega=0$ peak in the optical conductivity (left panels). This peak reflects the coherent tunneling of the electronic carrier as the TLS states become long-range entangled with  the electronic motion,
therefore suppressing low-energy scattering processes \cite{depolarone,optcondDMFT}. The spectra in panel (c) reveal that the emergence of this coherent peak coincides with the same condition $T\lesssim \omega_0$ inferred from the resistivity (see also Fig. \ref{fig:spectra-vsw0} below). 
A second phenomenon occurs if we increase $\omega_0$ further: as shown in panel (d), the full suppression of localization corrections inferred from the resistivity (panel h) corresponds to a fundamental change of shape for the optical absorption peak.

\begin{figure}[h]
        \includegraphics[width=8cm]{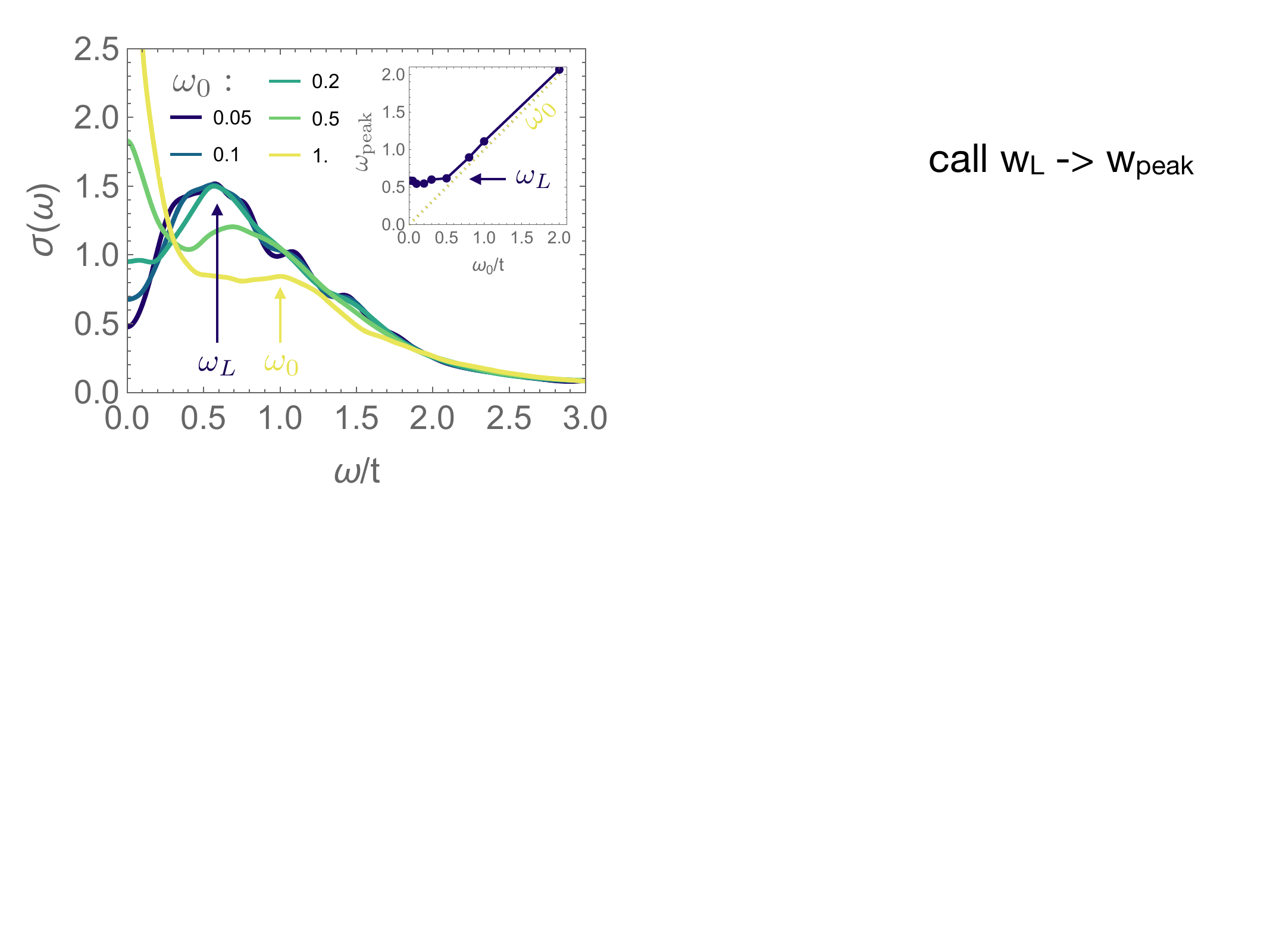}
\caption{Evolution of the optical conductivity with $\omega_0$, for $g/t=1$ and $T/t=1$. The inset shows the position of the finite-frequency peak as a function of $\omega_0$, crossing over from a localization peak at $\omega_L$ to a more conventional electron-TLS scattering resonance at $\omega_0$. }
    \label{fig:spectra-vsw0}
\end{figure}

To track the origin of this change of behavior, we show in Fig. \ref{fig:spectra-vsw0} the optical spectra at a fixed temperature $T/t=1$, varying instead the dynamical scale $\omega_0$ with respect to it (labels). First of all, we see that the $\omega=0$ peak appears when $\omega_0\gtrsim T$, confirming the emergence of low-temperature coherent transport  discussed previously. Second, the localization-induced DDP  is washed out when $\omega_0$ reaches the localization peak frequency $\omega_L$ (arrows and inset of Fig. \ref{fig:spectra-vsw0}). At this point, the fluctuations of the TLS are no longer seen by the electron as a slow (quasi-static) potential, and therefore they lose their ability to localize the electronic wavefunction: the quantum interferences at the origin of localization cannot be sustained if the very constituents of the  potential wells responsible for localization (here the emergent disorder arising from the fluctuating TLS, with a timescale set by $\omega_0$) move faster than the time it takes for the electronic wavefunction to localize (set by $\omega_L$). 

As localization corrections are washed out for $\omega_0>\omega_L$, the localization-induced DDP evolves into a more conventional resonance originating from individual electron-TLS scattering events. The threshold for these processes is  located   at $\omega \simeq \omega_0$ (arrow and inset of  Fig. \ref{fig:spectra-vsw0}).

Some remarks are in order here. First, the results presented here demonstrate that sufficiently slow \textit{quantum} fluctuations are able to sustain transient (Anderson) localization. This extends previous theoretical realizations of transient localization, that have relied instead  either on the interaction with classical degrees of freedom \cite{Fratini-AdvMat16,SciPost} or with quantum degrees of freedom in the classical regime $T>\omega_0$ \cite{rammalPRL2024}. 

Second, and directly related to the previous point, the existence of two crossover temperatures, $T\sim \omega_0$ for the establishment of coherent motion and $T\sim \omega_L$ for the  complete disappearance of localization effects, implies that in the enclosed temperature interval the localization-induced displaced Drude peak coexists with a more conventional Drude peak centered at $\omega=0$. This situation can be seen in both Fig. \ref{fig:spectra-vsT}(c) and \ref{fig:spectra-vsw0}. A regime with two coexisting peaks was also reported recently in the Holstein electron-boson interaction model \cite{MitricPRB2025,JankovicPRB24}.

\subsection{Frequency-dependence of vertex corrections}

\begin{figure}[h]
        \includegraphics[width=8cm]{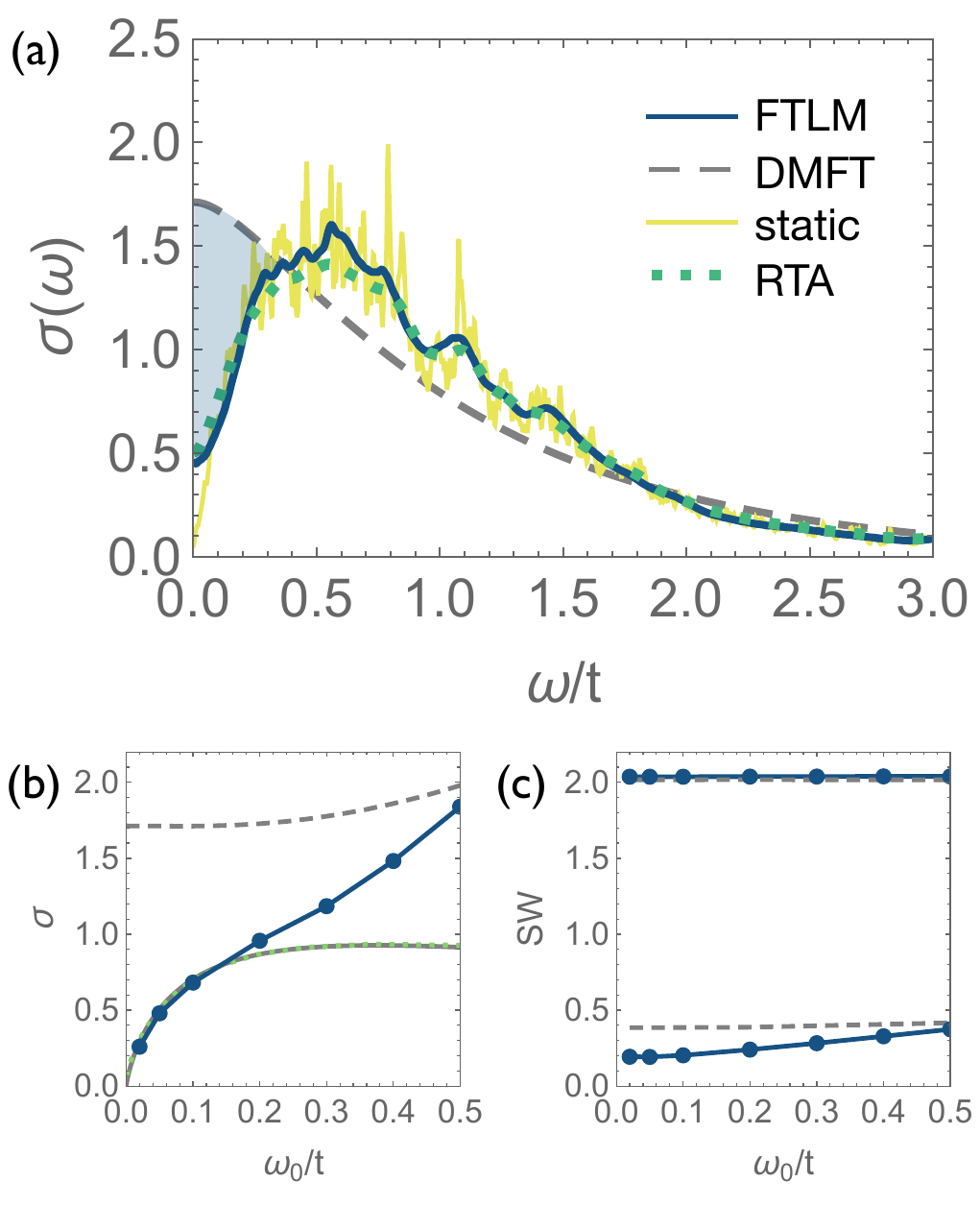}
\caption{(a) Optical conductivity calculated from FTLM for $g/t=1$, $\omega_0/t=0.05$, $T/t=1$ for $N_s=16$ with periodic boundary conditions, with $M=1200$ Lanczos iterations and $N_r=100$  realizations (blue solid line). Also shown are the results from DMFT (gray dashed), static TLS limit (yellow) and relaxation time approximation (green dotted).
(b)  d.c. conductivity at $T=1$ shown as a function of $\omega_0$. The gaussian filter is $\delta/t=0.04$ for all curves except at the lowest $\omega_0/t=0.02$, where we have taken $\delta/t=0.01$ (same legends as in panel (a)). The underlying gray line is the analytical Eq. (\ref{eq:AnalyticRTA}) with $\bar\omega=0.375 t$ (c) Evolution of the partial and total spectral weights with $\omega_0$.
}
    \label{fig:spectra-RTA}
\end{figure}

There exist two different types of vertex corrections to the conductivity \cite{Mahan}: (i) The quantum corrections originating from non-local interference processes, that are at the origin of localization, already discussed and (ii) Interaction processes that account for the geometrical aspects of scattering and that are responsible for the difference between the transport scattering rate and the quasiparticle scattering rate  (the former being dominantly affected by backward scattrering). The agreement between the FTLM and DMFT resistivity at $\omega_0 = \omega_L$, i.e. in a regime where localization effects play no role, indicates that in the present local electron-TLS scattering problem vertex corrections of type (ii) are irrelevant, and that the transport and quasiparticle scattering rates coincide.
This agrees with the conclusions drawn in Ref. \cite{Fratini20} for local electron-boson interactions in dimensions $d=1$.

Fig. \ref{fig:spectra-RTA} shows a representative optical conductivity spectrum calculated both from FTLM and DMFT at $T=1$, $g=1$ and a low $\omega_0/t=0.05$, highlighting how localization vertex corrections of type (i) depend on frequency. Similar to the electron-boson problem \cite{rammalPRL2024}, the comparison shows that full agreement between the two methods is recovered at high frequencies, $\omega \gtrsim \omega_L$, even when localization corrections are evident in the DDP range $\omega_0 \lesssim \omega\lesssim \omega_L$ and in the resistivity. 
To quantify these considerations, in Fig. \ref{fig:spectra-RTA}(b) we report the evolution of the d.c. conductivity as a function of $\omega_0$, and in Fig. \ref{fig:spectra-RTA}(c) the partial ($\Lambda=0.25 t$) and total ($\Lambda=\infty$) integrated optical spectral weights, defined as 
\begin{equation}
\mathrm{SW}(\Lambda)= \int_0^\Lambda d\omega \sigma(\omega).
\end{equation}

The FTLM conductivity is always lower than the DMFT result (a), and so is the partial spectral weight at low frequencies (b). The total spectral weight is instead precisely captured by DMFT: this means that localization corrections are responsible for a redistribution of spectral weight, without affecting the total integrated absorption. Since the latter is proportional to the total kinetic energy  \cite{Maldague}, this implies that the kinetic energy is itself precisely captured by DMFT, being entirely determined by one-particle properties and insensitive to non-local corrections.  The kinetic energy appears to be independent on $\omega_0$ in the explored window of Rabi frequencies $\omega_0 \lesssim T$.

\section{Relaxation time approximation}
We are now in a position to reexamine the relaxation time approximation (RTA),  assessing how it performs quantitatively in comparison to the exact numerical results obtained above. 

The RTA is a conceptually simple approach that has been traditionally used to construct the transient localization scenario \cite{Fratini-AdvMat16,PRBR11}. It builds on a reference disordered system exhibiting full Anderson localization, and includes the effect of slow fluctuations by applying an exponential relaxation to the current-current correlation function \cite{PRBR11,Fratini-AdvMat16,Dynamicallocalization},  mimicking the decay of quantum interferences when the disordered landscape is dynamic.
In practice one starts by evaluating the full optical conductivity of a reference system with static disorder, $\sigma_{static}(\omega)$, corresponding to Eq. (\ref{eq:TLSNodisorder}) with $\omega_0=0$. This step is numerically straightforward due to the non-interacting nature of the  problem. The effects of a finite $\omega_0$ are then approximately included by performing the following Lorentzian convolution: 
\begin{eqnarray}
& & \sigma^{(RTA)}(\omega,T) =   \frac{1}{\pi} \frac{\tanh{(\omega/2T)}}{\omega} \times  \label{eq:RTAw} \\
& & \times \int_{-\infty}^{\infty} \frac{\omega'}{\tanh{(\omega'/2T)}} \frac{p }{(\omega - \omega')^2 + p^2} \ \sigma_{static}(\omega')\, d\omega'.
\nonumber 
\end{eqnarray}
with $p$ a relaxation rate of the order of the fluctuation rate of dynamical disorder, here the Rabi frequency $\omega_0$. 
The d.c. conductivity is obtained  the zero-frequency limit of the previous expression,
\begin{equation}
\sigma^{(RTA)}_{d.c.} (T)=   \frac{p}{\pi T} \int_{0}^{\infty}  \frac{\omega}{\tanh{(\omega/2T)}} \frac{1}{\omega^2 + p^2} \sigma_{static}(\omega)\, d\omega, \label{eq:RTA}
\end{equation}
and the resistivity is $\rho^{(RTA)}=1/\sigma^{(RTA)}_{d.c.}$.
The results for the resistivity vs $T$, d.c. conductivity vs $\omega_0$  and optical conductivity vs $\omega$ obtained from Eqs. (\ref{eq:RTAw}) and (\ref{eq:RTA}) are shown as green dotted lines in Fig. \ref{fig:spectra-vsT} (e-h), Fig. \ref{fig:spectra-RTA}(b) and Fig. \ref{fig:spectra-RTA}(a) respectively. In all cases we have evaluated the RTA formulas by setting $p=\omega_0$, as this value provides the best agreement with the FTLM data (a lower $p\simeq 0.45\omega_0$ was found in the Holstein model \cite{rammalPRL2024}).

Fig. \ref{fig:spectra-vsT} shows that for all $\omega_0\lesssim \omega_L$ the exact temperature dependence of the resistivity is captured by the RTA with high accuracy, except for the weakly activated behavior present at very low $\omega_0$ and $T\lesssim t$, cf. panels (e-f). 
As this upturn is only seen  in the FTLM data, we
tentatively it to the formation of bond polarons, a phenomenon that involves a self-consistent polarization of the TLS degrees of freedom beyond RTA, as well as the inclusion of non-local processes beyond single-site DMFT.
Other than this, the RTA quantitatively reproduces not only the temperature dependence, but also the $\omega_0$-dependence of the conductivity at sufficiently low $\omega_0$, as illustrated in Fig. \ref{fig:spectra-RTA}(b). Fig. \ref{fig:spectra-RTA}(a) shows that, in fact, the entire frequency dependence of the optical conductivity is very accurately described by the RTA, including some of the fine structure at finite frequency.

We conclude this Section by deriving an  analytical formula for the d.c. conductivity, based on the RTA Eq. (\ref{eq:RTA}), that closely reproduces the numerical data. For this we assume that the optical conductivity in the static limit is of the form $\sigma_{static}(\omega)=(\pi/2)A(T) |\omega| /(1+\omega^2/\bar\omega^2)$, describing a vanishing d.c. conductivity and a finite-frequency peak at $\bar\omega \simeq \omega_L$ as the one shown in Fig. \ref{fig:spectra-RTA}(a) (yellow), and whose temperature dependence is contained in the prefactor $A(T)$ (the precise behaviors for $\omega\to 0$ and at $\omega >\bar\omega$ are irrelevant to the final result). Evaluating the RTA Eq. (\ref{eq:RTA}) on this toy model yields, for $T\gtrsim \bar\omega \gtrsim \omega_0$, 
\begin{equation}
    \sigma^{(RTA)}_{d.c.}(T)\simeq A(T) \ \omega_0  \frac{\log(\bar\omega/\omega_0)}{1-(\omega_0/\bar\omega)^2}.
    \label{eq:AnalyticRTA}
\end{equation}
Eq. (\ref{eq:AnalyticRTA}) shows that the zeroth order RTA result $\sigma \propto \omega_0$ \cite{Fratini-AdvMat16,PRBR11} only holds at very low values of $\omega_0$, while it is strongly renormalized by the presence of the DDP frequency cutoff already at moderate values of $\omega_0$.
The result obtained after optimizing the value of $\bar\omega$ is reported as a gray line in Fig. \ref{fig:spectra-RTA}(b), and it is essentially indistinguishable from the full RTA result  (green dotted), and the numerical FTLM result itself (full line and markers).

\section{Results in the fast TLS regime}

\begin{figure}
    \centering
    \includegraphics[width=8cm]{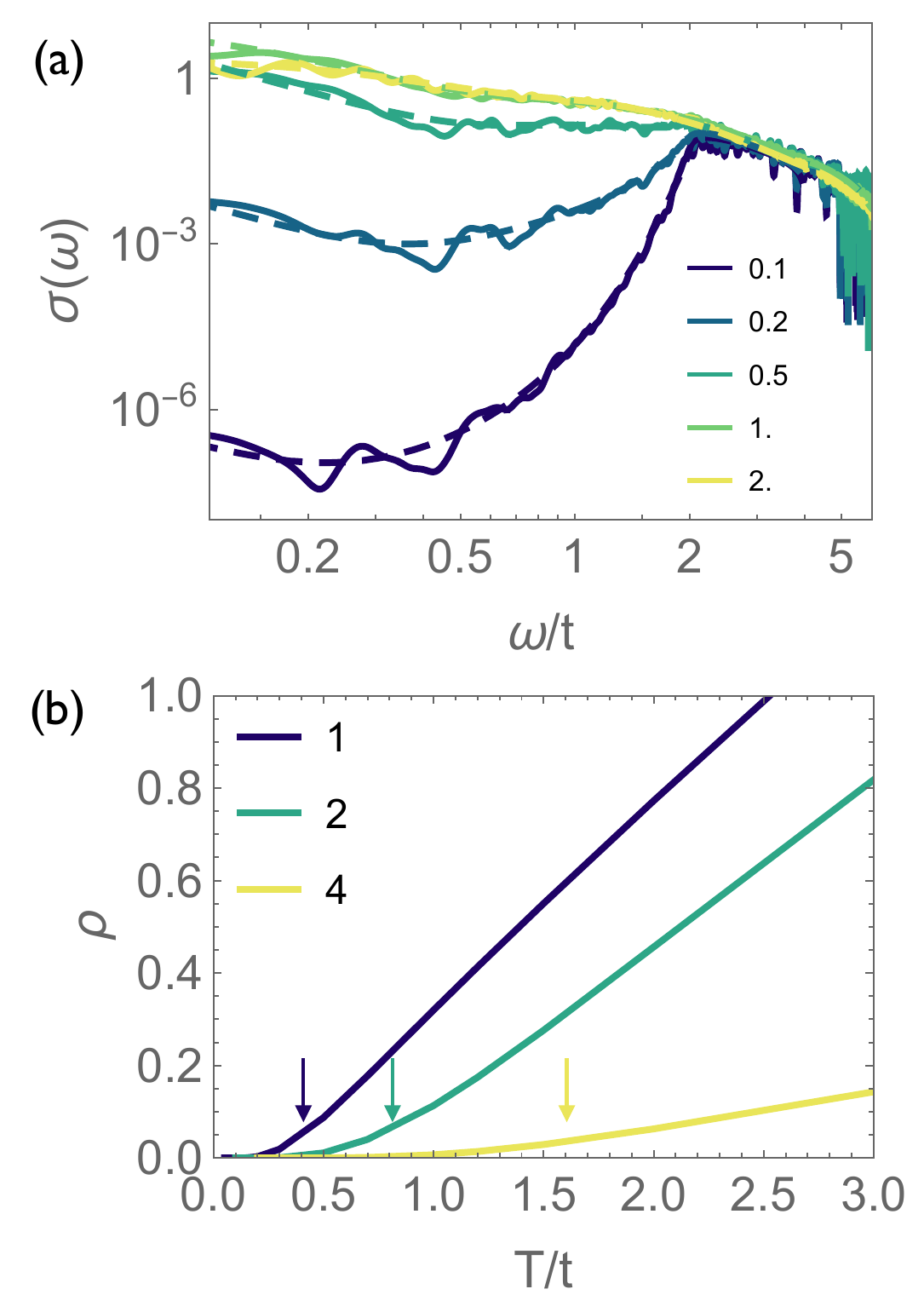}
    \caption{(a) Optical conductivity at $\omega_0/t=2$ and different temperatures (legends) comparing FTLM on a chain of length $N_s=18$ with $M=600$ Lanczos iterations and averaged over $500$ twisted boundary conditions  (solid lines, Gaussian filter $\delta=0.02t$) and DMFT (dashed lines). (b) DMFT resistivity vs temperature at different $\omega_0/t$ (legends). The arrows mark the temperature $T^*=0.4 \omega_0$.}
    \label{fig:largew0}
\end{figure}
In the preceding Sections we have shown that, due to the suppression of vertex corrections, the DMFT results match the FTLM results   
at the largest explored $\omega_0 \simeq \omega_L \simeq 0.5 t$. Beyond this point, the FTLM method with periodic boundary conditions reaches its limits of applicability for the limited cluster sizes attainable. This is signaled by strong finite size effects on both the optical conductivity and the extrapolated d.c. value persisting even at the largest attainable size $N_s=18$, indicating that the results have not reached the thermodynamic limit. This happens because the localization effects that were responsible for the fast convergence of the method are lost when the TLS are too fast, as discussed above. 

The large $\omega_0$ regime can still be addressed numerically, but one needs to treat finite-size effects more carefully. This can be achieved through twisted boundary condition averaging as was done in the case of electron-boson interactions in Ref. \cite{rammalPRL2024,TBC_1}.
Fig. \ref{fig:largew0}(a) shows the optical conductivity spectra obtained at different temperatures upon averaging over $500$ twisted boundary conditions for a  value $\omega_0/t=2$, much larger than those shown in Fig. \ref{fig:spectra-vsT} (full lines). The optical spectra show a temperature independent  threshold at $\omega=\omega_0$ corresponding to the excitation of one TLS from the lower to the upper state, and a  thermally activated absorption below this threshold.

The comparison with the DMFT results shows a striking agreement at all temperatures, all frequencies, and over more than 7 orders of magnitude of conductivity. This demonstrates that the handshake between FTLM and DMFT,  shown in Figs. \ref{fig:spectra-vsT}(d,h) at $\omega_0 \simeq \omega_L$,  persists in the entire large $\omega_0$ regime. More generally, Fig. \ref{fig:largew0}(a) provides a spectacular benchmark of the twisted boundary condition averaging technique, showing that it allows to obtain converged spectra reaching the thermodynamic limit at finite frequencies, even when localization effects are not there to confine the carrier wavefunction within the bounds of the cluster size.

Because of the good agreement demonstrated here between DMFT and FTLM, we can now use DMFT to calculate the d.c. resistivity in the fast TLS regime. This is advantageous as this method works directly at the thermodynamic limit,  being by construction devoid of finite-size effects down to $\omega=0$  (see Appendix \ref{app:DMFT}). Reaching the d.c. limit in FTLM is instead extremely challenging, as the width of the coherent peak at $\omega=0$ becomes exponentially narrow for $T \lesssim \omega_0$, rapidly falling below the numerical resolution. 

We report in  Fig. \ref{fig:largew0}(b) the temperature dependence of the resistivity calculated for increasing values of $\omega_0$. It features an exponential suppression at low temperatures, followed by an approximately linear increase at larger temperatures. The crossover temperature is approximately located at  $T\simeq 0.4 \omega_0$, indicated by arrows. For the moderate coupling strengths $g/\omega_0 \lesssim 1$ explored here, both the optical and d.c. conductivity in the fast fluctuation regime are reminiscent of the picture arising in the Holstein model  \cite{rhopolaronPRL03,optcondDMFT}. 

Although a study of the large $g/\omega_0$ regime at large $\omega_0$ goes beyond the scope of this work, we mention here that a fundamental difference between the two models should arise in this case: on-site polaron formation, that characterizes the strong coupling regime of the Holstein model, will be strongly suppressed in the electron-TLS interaction model as it involves several oscillator states that are by construction not available  (see also Appendix \ref{app:DMFT}).

\section{Concluding remarks}

In this paper, we have studied the transport properties of a charge carrier interacting with two-level systems located at all lattice sites. The transport scenarios that emerge from our study depend strongly on the  timescale of the TLSs relative to the free charge dynamics, characterized by the ratio $\omega_0/t$, as well as to the buildup time for localization by the emergent dynamical disorder, characterized by the ratio $\omega_0/\omega_L$.

Our results demonstrate that in the slow TLS regime, $\omega_0 \lesssim  t$, the transient localization scenario  originally established for interactions with slow  bosons \cite{AdvMat16,rammalPRL2024} also applies to electronic carriers coupled with TLS. We conclude that 
the key ingredient in the transient localization mechanism is the slowness of the quantum fluctuations, rather than their bosonic nature, which makes this phenomenon more general than what was previously understood. In this slow TLS regime, the relaxation-time approximation,  that incorporates Anderson localization corrections at the dynamical level at a very modest computational cost, is found to broadly reproduce the exact numerical data. 

Localization corrections are found to be progressively suppressed upon increasing the dynamical scale $\omega_0$, until both the d.c. and dynamic conductivities recover the picture expected in the presence of independent scattering processes. In this regime the results obtained by FTLM converge to the results of DMFT, indicating that vertex corrections become negligible.

Besides constituting an interesting problem \textit{per se}, the numerical advantage brought by the reduced Hilbert space of the electron-TLS interaction model could enable accurate numerical studies of more complex questions, that are otherwise out of reach in the electron-boson case. For example, the theoretical  framework developed here could find direct application to the study of the interplay between elastic scattering by quenched disorder and inelastic scattering by quantum degrees of freedom  
\cite{GirvinJonsonPRB80,gogolin75,gogolin77},
a problem for which exact numerical results simply do not exist. Similarly, one could study the emergence of thermally activated conduction in disordered insulators, going beyond the weak inelastic scattering assumption that is the foundation of all the theories available today --- ranging from Arrhenius \cite{Miller-Abrahams} to variable range hopping \cite{MottKaveh} to Efros-Shklovskii Coulomb gap law \cite{Efros}.
On a longer term, a more ambitious program would be to incorporate many-body physics at finite carrier concentration, exploring how electron-electron and interactions affect the coupling to dynamical degrees of freedom, and vice-versa.

\acknowledgments
S.C. acknowledges funding from  NextGenerationEU National Innovation Ecosystem grant  PE00000023 - IEXSMA - CUP E63C22002180006.

\bibliographystyle{unsrt}  
\bibliography{TLSTLFTLM}

\begin{thebibliography}{10}

\bibitem{AdvMat16}
S.~Fratini, D.~Mayou, and S.~Ciuchi.
\newblock The transient localization scenario for charge transport in
  crystalline organic materials.
\newblock {\em Adv. Funct. Mater}, 26:2292, 2016.

\bibitem{Hussey}
N.~Hussey, K.~Takenaka, and H.~Takagi.
\newblock Universality of the {Mott–Ioffe–Regel} limit in metals.
\newblock {\em Philos Mag.}, 84:2847, 2004.

\bibitem{PhillipsScience22}
Philip~W. Phillips, Nigel~E. Hussey, and Peter Abbamonte.
\newblock Stranger than metals.
\newblock {\em Science}, 377(6602):eabh4273, 2022.

\bibitem{Troisi06}
A.~Troisi and G.~Orlandi.
\newblock Charge-transport regime of crystalline organic semiconductors:
  Diffusion limited by thermal off-diagonal electronic disorder.
\newblock {\em Phys. Rev. Lett.}, 96:086601, 2006.

\bibitem{PRBR11}
S.~Ciuchi, S.~Fratini, and D.~Mayou.
\newblock Transient localization in crystalline organic semiconductors.
\newblock {\em Phys. Rev. B}, 83:081202(R), 2011.

\bibitem{rammalPRL2024}
Hadi Rammal, Arnaud Ralko, Sergio Ciuchi, and Simone Fratini.
\newblock Transient localization from the interaction with quantum bosons.
\newblock {\em Phys. Rev. Lett.}, 132:266502, Jun 2024.

\bibitem{SciPost}
S.~Fratini and S.~Ciuchi.
\newblock {Displaced Drude peak and bad metal from the interaction with slow
  fluctuations.}
\newblock {\em SciPost Phys.}, 11:39, 2021.

\bibitem{HellerPRL}
Joonas Keski-Rahkonen, Xiaoyu Ouyang, Shaobing Yuan, Anton~M. Graf, Alhun
  Aydin, and Eric~J. Heller.
\newblock Quantum-acoustical drude peak shift.
\newblock {\em Phys. Rev. Lett.}, 132:186303, May 2024.

\bibitem{FTLM_1}
J.~Jaklič and P.~Prelovšek.
\newblock Lanczos method for the calculation of finite-temperature quantities
  in correlated systems.
\newblock {\em Phys. Rev. B}, 49, 1994.

\bibitem{prelovsek}
P.~Prelov{\v{s}}ek and J.~Bon{\v{c}}a.
\newblock {\em Ground State and Finite Temperature Lanczos Methods}, pages
  1--30.
\newblock Springer Berlin Heidelberg, Berlin, Heidelberg, 2013.

\bibitem{TLSJEORGE1}
Noga Bashan, Evyatar Tulipman, J\"org Schmalian, and Erez Berg.
\newblock Tunable non-fermi liquid phase from coupling to two-level systems.
\newblock {\em Phys. Rev. Lett.}, 132:236501, Jun 2024.

\bibitem{TLSJEORGE2}
Evyatar Tulipman, Noga Bashan, J\"org Schmalian, and Erez Berg.
\newblock Solvable models of two-level systems coupled to itinerant electrons:
  Robust non-fermi liquid and quantum critical pairing.
\newblock {\em Phys. Rev. B}, 110:155118, Oct 2024.

\bibitem{PRLIntrinsicQuantumGlasses}
Vassiliy Lubchenko and Peter~G. Wolynes.
\newblock Intrinsic quantum excitations of low temperature glasses.
\newblock {\em Phys. Rev. Lett.}, 87:195901, Oct 2001.

\bibitem{AndersonPhilMag72}
P.~w.~Anderson, B.~I. Halperin, and c.~M.~Varma~and.
\newblock Anomalous low-temperature thermal properties of glasses and spin
  glasses.
\newblock {\em The Philosophical Magazine: A Journal of Theoretical
  Experimental and Applied Physics}, 25(1):1--9, 1972.

\bibitem{Bashan25}
Noga Bashan, Evyatar Tulipman, J\"org Schmalian, and Erez Berg.
\newblock Tunable non-fermi liquid phase from coupling to two-level systems.
\newblock {\em Phys. Rev. Lett.}, 132:236501, Jun 2024.

\bibitem{CaponeCartaPRB06}
M.~Capone, P.~Carta, and S.~Ciuchi.
\newblock Dynamical mean field theory of polarons and bipolarons in the
  half-filled {Holstein} model.
\newblock {\em Phys. Rev. B}, 74, 1996.

\bibitem{depolarone}
S.~Ciuchi, F.~de~Pasquale, S.~Fratini, and D.~Feinberg.
\newblock Dynamical mean-field theory of the small polaron.
\newblock {\em Phys. Rev. B}, 56:4494--4512, Aug 1997.

\bibitem{PRBStatisticaltheory}
Subroto Mukerjee, Vadim Oganesyan, and David Huse.
\newblock Statistical theory of transport by strongly interacting lattice
  fermions.
\newblock {\em Phys. Rev. B}, 73:035113, Jan 2006.

\bibitem{auerbach}
Netanel~H. Lindner and Assa Auerbach.
\newblock Conductivity of hard core bosons: A paradigm of a bad metal.
\newblock {\em Phys. Rev. B}, 81:054512, Feb 2010.

\bibitem{spin-Holstein}
Sergio Ciuchi and Simone Fratini.
\newblock Strange metal behavior from incoherent carriers scattered by local
  moments.
\newblock {\em Phys. Rev. B}, 108:235173, Dec 2023.

\bibitem{optcondDMFT}
S.~Fratini and S.~Ciuchi.
\newblock Optical properties of small polarons from dynamical mean-field
  theory.
\newblock {\em Phys. Rev. B}, 74, 2006.

\bibitem{Fratini-AdvMat16}
S.~Fratini, D.~Mayou, and S.~Ciuchi.
\newblock The transient localization scenario for charge transport in
  crystalline organic materials.
\newblock {\em Adv. Funct. Mater}, 26:2292, 2016.

\bibitem{MitricPRB2025}
P.~Mitri\ifmmode~\acute{c}\else \'{c}\fi{},
  V.~Dobrosavljevi\ifmmode~\acute{c}\else \'{c}\fi{}, and
  D.~Tanaskovi\ifmmode~\acute{c}\else \'{c}\fi{}.
\newblock Precursors to anderson localization in the holstein model: Quantum
  and quantum-classical solutions.
\newblock {\em Phys. Rev. B}, 111:L161105, Apr 2025.

\bibitem{JankovicPRB24}
Veljko Jankovi\ifmmode~\acute{c}\else \'{c}\fi{}, Petar
  Mitri\ifmmode~\acute{c}\else \'{c}\fi{}, Darko
  Tanaskovi\ifmmode~\acute{c}\else \'{c}\fi{}, and Nenad
  Vukmirovi\ifmmode~\acute{c}\else \'{c}\fi{}.
\newblock Vertex corrections to conductivity in the holstein model: A
  numerical-analytical study.
\newblock {\em Phys. Rev. B}, 109:214312, Jun 2024.

\bibitem{Mahan}
G.~D. Mahan.
\newblock {\em Many-Particle Physics}.
\newblock Kluwer Academic/Plenum Publisher, New Yok, 2000.

\bibitem{Fratini20}
S.~Fratini and S.~Ciuchi.
\newblock Dynamical localization corrections to band transport.
\newblock {\em Phys. Rev. Research}, 2:013001, 2020.

\bibitem{Maldague}
Pierre~F. Maldague.
\newblock Optical spectrum of a hubbard chain.
\newblock {\em Phys. Rev. B}, 16:2437--2446, Sep 1977.

\bibitem{Dynamicallocalization}
S.~Fratini and S.~Ciuchi.
\newblock Dynamical localization corrections to band transport.
\newblock {\em Phys. Rev. Res.}, 2:013001, Jan 2020.

\bibitem{TBC_1}
Didier Poilblanc.
\newblock Twisted boundary conditions in cluster calculations of the optical
  conductivity in two-dimensional lattice models.
\newblock {\em Phys. Rev. B}, 44, 1991.

\bibitem{rhopolaronPRL03}
S.~Fratini and S.~Ciuchi.
\newblock Dynamical mean-field theory of transport of small polarons.
\newblock {\em Phys. Rev. Lett.}, 91:256403, Dec 2003.

\bibitem{GirvinJonsonPRB80}
S.~M. Girvin and M.~Jonson.
\newblock Dynamical electron-phonon interaction and conductivity in strongly
  disordered metal alloys.
\newblock {\em Phys. Rev. B}, 22:3583--3597, Oct 1980.

\bibitem{gogolin75}
AA~Gogolin, VI~Mel'Nikov, and EI~Rashba.
\newblock Conductivity in a disordered one-dimensional system induced by
  electron-phonon interaction.
\newblock {\em Soviet Journal of Experimental and Theoretical Physics}, 42:168,
  1975.

\bibitem{gogolin77}
AA~Gogolin, VI~Mel'Nikov, and E{\'e}~I Rashba.
\newblock Effect of dispersionless phonons on the kinetics of electrons in
  one-dimensional conductors.
\newblock {\em Soviet Journal of Experimental and Theoretical Physics}, 45:330,
  1977.

\bibitem{Miller-Abrahams}
Allen Miller and Elihu Abrahams.
\newblock Impurity conduction at low concentrations.
\newblock {\em Physical Review}, 120(3):745, 1960.

\bibitem{MottKaveh}
N.~F. Mott and M.~Kaveh.
\newblock Metal-insulator transitions in non-crystalline systems.
\newblock {\em Adv. Phys.}, 34:329, 1985.

\bibitem{Efros}
Al~L {\'E}fros and Boris~I Shklovskii.
\newblock Coulomb gap and low temperature conductivity of disordered systems.
\newblock {\em Journal of Physics C: Solid State Physics}, 8(4):L49, 1975.

\bibitem{DMFTreview}
Antoine Georges, Gabriel Kotliar, Werner Krauth, and Marcelo~J. Rozenberg.
\newblock Dynamical mean-field theory of strongly correlated fermion systems
  and the limit of infinite dimensions.
\newblock {\em Rev. Mod. Phys.}, 68:13--125, Jan 1996.

\bibitem{optcondpolaronPRB06}
S.~Fratini and S.~Ciuchi.
\newblock Optical properties of small polarons from dynamical mean-field
  theory.
\newblock {\em Phys. Rev. B}, 74:075101, Aug 2006.

\end{thebibliography}


\appendix

\section{DMFT of the TLS model}
\label{app:DMFT}

\subsection{DMFT solution}
The expression of the retarded Green's function for the model Eq. (\ref{eq:TLSNodisorder}) can be obtained using dynamical mean-field theory (DMFT) using a derivation analogous to that used for the Holstein model  \cite{depolarone}. Shifting the site energy by the constant amount $\omega_0/2$ allows to rewrite the energy splitting term $\frac{\omega_0}{2}\sum _i \sigma^z_i$ as $\omega_0\sum_i n_i$, where $n_i=0,1$. The model  can then straightforwardly be mapped onto an equivalent single impurity problem 
\begin{equation}
    H_{imp} = \omega_0 n      - g f^{\dagger}f\sigma^x + H_{hyb} 
\label{eq:HamAIM}
\end{equation}
where $H_{hyb}$ describes a fermionic bath hybridized with the impurity electron $f$  \cite{DMFTreview} and $n=0,1$.
For one electron the impurity Green function $G=-i\langle Tf(t)f^\dagger(0)\rangle$ in the frequency domain can be calculated using the resolvent
\begin{equation}
    G(\omega)=\frac{1}{Z}\sum^{1}_{n=0} e^{-\beta n \omega_0} G_{n,n}(\omega+n\omega_0)
\end{equation}
with 
\begin{equation}
G_{n,n}(\omega) = \langle n,0|f \frac{1}{\omega-H_{imp}} f^\dagger|n,0\rangle.
\end{equation}
Here $Z=\sum^{1}_{n=0} e^{-\beta n \omega_0}$  is the TLS partition function and $|n,0\rangle$ is the product state of the vacuum for the electron and the TLS in the state $n=0,1$. Using the operator identity $(\omega-H_0-V)^{-1}=(\omega-H_0)^{-1}+(\omega-H_0)^{-1}V(\omega-H_0-V)^{-1}$ with $V=-g f^{\dagger}f\sigma^x$, the matrix $\mathbf{G}$ in the TLS components can be obtained through the solution of:
\begin{equation}
\mathbf{G}=\mathbf{G_{0}}-g\mathbf{G_{0}}\sigma^x \mathbf{G},
\label{eq:resolvent}
\end{equation}
where $\mathbf{G_{0}}$ is a diagonal matrix in the TLS index $n$,
\begin{equation}
G^{0}_{n,n}(\omega) = \langle 0, n|f \frac{1}{\omega-n\omega_0-H_{hyb}} f^\dagger|n,0\rangle.
\label{eq:G0}
\end{equation}
By inverting Eq. (\ref{eq:resolvent}) we obtain the Green's function
\begin{eqnarray}
    G(\omega)&=&\frac{1}{1+e^{-\beta \omega_0}}\left [ \frac{1}{G^{-1}_0(\omega)-g^2 G_0(\omega-\omega_0)} \right. \nonumber \\
    &+& \left.\frac{e^{-\beta\omega_0}}{G^{-1}_0(\omega)-g^2 G_0(\omega+\omega_0)}\right ].
    \label{eq:Green}
\end{eqnarray}
Here $G_0(\omega)$ is given by Eq. (\ref{eq:G0}) with $n=0$  and is actually computed by enforcing the DMFT self-consistency condition \cite{DMFTreview}.
In practice, Eq. (\ref{eq:Green}) can be  obtained directly from the DMFT solution of the Holstein model (Eqs. 39-42 in Ref. \cite{depolarone}) by truncating the continued fraction expansion to the first level, i.e. considering a phonon Hilbert space that includes $n=0,1$ only.

\begin{figure}[h]
\includegraphics[width=9cm]{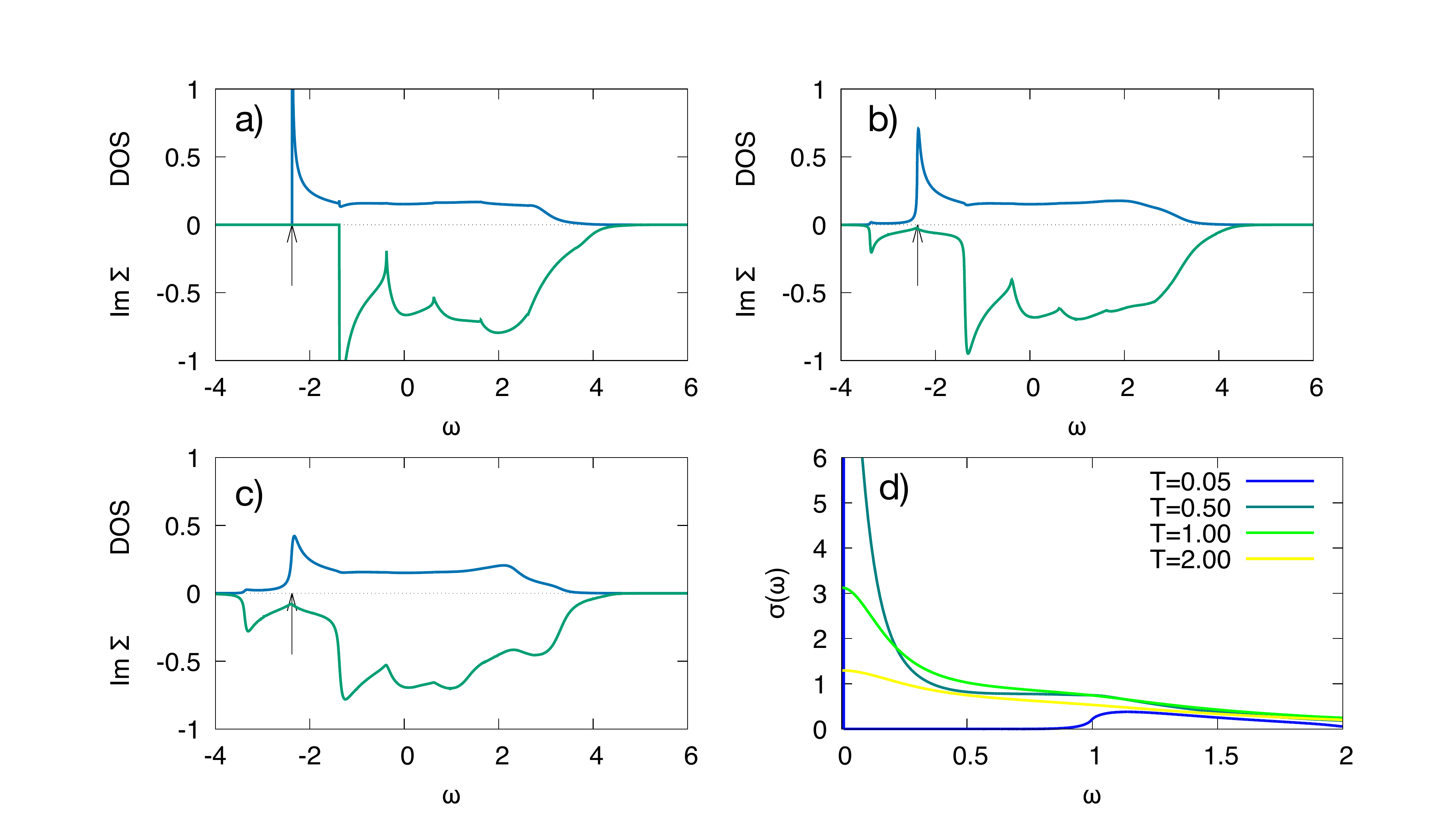}
\caption{a,b,c) DMFT DOS (blue) and imaginary part of the self-energy (green)  $g=1,\omega_0=1$ for a) $T=0$, b) $T=0.5$ c) $T=1$ (all in units of the transfer integral $t$). The arrows mark the ground state energy. d) Optical conductivity at different temperatures (legends) for the same values of $g,\omega_0$. }
    \label{fig:DMFT}
\end{figure}

Once that the Green's function $G$ and the self-energy $\Sigma(\omega)=G_0^{-1}(\omega)-G^{-1}(\omega)$ are known, the calculation of the optical conductivity and of the resistivity proceeds by evaluating the Kubo formula for the current-current correlation function in the absence of vertex corrections, as presented for example in Refs. \cite{rhopolaronPRL03,optcondpolaronPRB06}. Typical results for the renormalized density of states (DOS) $N^*(\omega)=-Im G(\omega)/\pi$ and the scattering rate $\Gamma(\omega)=-Im \Sigma(\omega)$  are depicted in Fig. \ref{fig:DMFT}. As in the Holstein polaron case, states at energies $E_0<\omega<E_0+\omega_0$ are fully coherent \cite{depolarone} at zero temperature, i.e. the scattering rate is identically zero in this energy interval, cf. Fig. \ref{fig:DMFT} (a). At non-zero temperature, these states progressively become incoherent due to thermal effects. In addition, thermally activated states appear in the spectral function at energies lower than the ground state energy, Figs. \ref{fig:DMFT} (b) and (c). These states are damped as demonstrated by the corresponding imaginary part of self-energy. As these states become populated in the DOS, the corresponding optical gap at $0<\omega<\omega_0$, separating the coherent Drude peak from the single TLS absorption band, is progressively filled, see Figs. \ref{fig:DMFT} (d) and \ref{fig:largew0}(a). 

In the slow TLS, high temperature regime $\omega_0 \ll t,\omega_0 \ll T$, Eq. (\ref{eq:Green})  can be simplified to
\begin{eqnarray}
    G(\omega) = & & \frac{1}{2}\left [ \frac{1}{G^{-1}_0(\omega)-g^2 G_0(\omega)} \right. \nonumber \\
    & & +\left. \frac{1}{G^{-1}_0(\omega)-g^2 G_0(\omega)}\right ].
    \label{eq:GreenAdiabatic}
\end{eqnarray}

This is equivalent to the Green's function of an electron moving on a lattice with binary disorder ($\pm g$) within the coherent potential approximation (CPA).

From Eq. (\ref{eq:Green}) we can obtain the self-energy as $G^{-1}=G^{-1}_0-\Sigma$. 
Denoting $\downarrow$ and $\uparrow$ respectively the ground state ($n=0$) and excited state ($n=1$) of the TLS, and introducing   the number of defects
\begin{equation}
n_i = \frac{\exp(-\omega_0/T)}{\exp(-\omega_0/T)+1}
\label{eq:impurities}
\end{equation}
that represents the number of thermally generated impurities (i.e. the number of TLS  in their excited state) 
we have
\begin{equation}
    \Sigma = \frac{\Sigma_{\downarrow}+n_i G^{-1}_0 \Delta }{1+n_i \Delta}
    \label{eq:Sigma}
\end{equation}
where 
\begin{equation}
    \Delta = \frac{\Sigma_{\downarrow}-\Sigma_{\uparrow}}{G^{-1}_0-\Sigma_{\uparrow}},
\end{equation}
$\Sigma_{\downarrow}(\omega)=g^2 G_0(\omega-\omega_0)$ and $\Sigma_{\uparrow}(\omega)=g^2 G_0(\omega+\omega_0)$.

In the low temperature limit $T\ll \omega_0$, $n_i$ is exponentially small and Eq. (\ref{eq:Sigma}) can be expanded to lowest order in $n_i$ yielding
\begin{equation}
    \Sigma \simeq \Sigma_{\downarrow} + n_i (\Sigma_{\uparrow}-\Sigma_{\downarrow}) \frac{G^{-1}_0-\Sigma_{\downarrow}}{G^{-1}_0-\Sigma_{\uparrow}}.
    \label{eq:SigmaLow}
\end{equation}
For weak coupling and for large negative frequencies the fraction appearing in Eq. (\ref{eq:SigmaLow}) can be approximated to 1 leading to 
$\Sigma = (1-n_i) \Sigma_{\downarrow}+n_i \Sigma_{\uparrow}$. The scattering rate $\Gamma=-Im \Sigma$ is given by
\begin{equation}
\Gamma(\omega) = -n_i g^2 Im G_0(\omega+\omega_0).
\label{eq:ScatteringRate}
\end{equation}

A form analogous to Eq. (\ref{eq:ScatteringRate}) could have been obtained formally within second order perturbation theory, in which case $G_0$ would be the non-interacting propagator.  The present derivation shows that Eq. (\ref{eq:ScatteringRate}) retains a non-perturbative character: its validity is enforced by the exponentially small number of thermal impurities $n_i$, and it is therefore not restricted to small values of the coupling constant $g$. Using the fact that for large negative frequencies $Im G_0(\omega-\omega_0)$ vanishes
around the band edges as $\omega^\alpha$ (in d=1 $\alpha=-1/2$) we have the following approximate result around the ground state energy:
\begin{equation}
\Gamma(\omega) \propto \exp(-\omega_0/T) \frac{g^2}{D} (\frac{\omega}{D})^\alpha.
\label{eq:ScatterinRate}
\end{equation}

\subsection{Transport properties}
\label{app:DMFTtransport}

The Kubo formula expresses the conductivity per particle (mobility $\mu$)
for a single electron as \cite{rhopolaronPRL03}
\begin{equation}
    \mu = \frac{\pi}{T} \frac{\int d\nu e^{-\beta \nu} {\cal D}(\nu)}{\int d\nu e^{-\beta \nu} {\cal N}(\nu)},
    \label{eq:Kubo}
\end{equation}
where
\begin{eqnarray}
    {\cal D}(\nu)&=&\int d\epsilon N(\epsilon)\phi(\epsilon) A^2(\nu,\epsilon)\label{eq:defD}\\ 
    {\cal N}(\nu)&=&\int d\epsilon N(\epsilon) A(\nu,\epsilon),  \label{eq:defN}
\end{eqnarray}
with $\phi(\epsilon)=4t^2-\epsilon ^2$ representing the squared band velocity and
$A$  the spectral function
\begin{equation}
    A(\nu,\epsilon)=\frac{\Gamma(\nu)}{\pi \left [(\nu-Re \Sigma(\nu)-\epsilon)^2+\Gamma^2(\nu) \right ]}.
    \label{eq:lorentzian}
\end{equation}   

For $\Gamma \ll t$ the integrals in $\epsilon$ appearing in Eqs. (\ref{eq:defN},\ref{eq:defD}) can be approximated as
\begin{eqnarray}
    {\cal D}(\nu)&=& N(\nu^*)\phi(\nu^*) \frac{1}{2 \pi \Gamma(\nu^*)}\label{eq:approxD}\\
    {\cal N}(\nu)&=&N(\nu^*)   \label{eq:approxN}
\end{eqnarray}
where $\nu^*=\nu-Re \Sigma(\nu)$, giving
\begin{equation}
    \mu = \frac{1}{2T} \frac{\int d\nu e^{-\beta \nu} N(\nu^*)\phi(\nu^*)/\Gamma(\nu^*)}{\int d\nu e^{-\beta \nu} N(\nu^*)}.
    \label{eq:mulowT}
\end{equation}
The previous results admit the following low temperature limit: if the scattering rate is small but finite ($\Gamma^*$) around the lower band edge ($\nu\equiv -D^*$), we can expand the integrands appearing in Eq. (\ref{eq:mulowT})   assuming that
${\cal N}(\nu^*)\simeq (\nu+D^*)^\alpha$ and therefore ${\cal N}(\nu^*)\phi(\nu^*)\simeq (\nu+D^*)^{\alpha+1}$. Performing the integrals yields 
\begin{equation}
    \mu = \frac{1}{2T \Gamma^*} c_\alpha,
    \label{eq:mulowlowT}
\end{equation}
where in $d=1$
$c_\alpha=(\alpha+1)  \ \lim_{\nu\rightarrow -2t} \phi(\nu)/(\nu+2t)$ with  $\alpha=-1/2$, yielding $c_{-1/2}=2t$.
From Eq. (\ref{eq:ScatterinRate}) we see that the leading temperature dependence for the mobility at low temperature is given by the thermal activation of impurities $n_i$, Eq. (\ref{eq:impurities}), independently of the relative values of $\omega_0$ and $t$.

\end{document}